\documentclass[a4paper,11pt]{article}
\usepackage{pos}

\usepackage{siunitx}
\usepackage{wrapfig}
\usepackage{placeins}
\usepackage{setspace}
\usepackage{amsmath}

\newcommand{\refeq}[1]{Eq.~(\ref{#1})}

\newcommand{\reffig}[1]{Fig.~\ref{#1}}

\newcommand{\refsec}[1]{Section~\ref{#1}}

\newcommand{\refref}[1]{Ref.~\cite{#1}}

\newcommand{\Nmu}{$N_\mu$}
\newcommand{\Enul}{$E_0$}
\newcommand{\Nav}{$\langle N_\mu \rangle$}

\newcommand{\Sone}{$S_{125}$}
\newcommand{\tta}{$\theta$}

\newcommand{\sibyllpre}{Sibyll~2.1}

\newcommand{\epos}{EPOS\nobreakdash-LHC}
\newcommand{\qgsjet}{QGSJet\nobreakdash-II.04}

\title{Multiplicity of TeV muons in air showers detected with IceTop and IceCube}

\manuallySeparateAuthors
\author*[a]{Stef Verpoest}
\author[\dag]{ for the IceCube collaboration}
\notes{\note{\protect\url{http://icecube.wisc.edu}}}

\affiliation[a]{Department of Physics and Astronomy, Ghent University,\\
9000 Gent, Belgium}

\emailAdd{stef.verpoest@icecube.wisc.edu}

\abstract{The IceCube Neutrino Observatory at the South Pole can provide unique tests of muon production models in extensive air showers by measuring both the low-energy (GeV) and high-energy (TeV) muon components. We present here a measurement of the TeV muon content in near-vertical air showers detected with IceTop in coincidence with IceCube. The primary cosmic-ray energy is estimated from the dominant electromagnetic component of the air shower observed at the surface. The high-energy muon content of the shower is studied based on the energy losses measured in the deep detector. Using a neural network, the primary energy and the multiplicity of TeV muons are estimated on an event-by-event basis. The baseline analysis determines the average multiplicity as a function of the primary energy between \SI{2.5}{\peta\eV} and \SI{250}{\peta\eV} using the hadronic interaction model \sibyllpre{}. Results obtained using simulations based on the post-LHC models \qgsjet{} and \epos{} are presented for primary energies up to \SI{100}{\peta\eV}. For all three hadronic interaction models, the measurements of the TeV muon content are consistent with the predictions assuming recent composition models. Comparing the results to measurements of GeV muons in air showers reveals a tension in the obtained composition interpretation based on the post-LHC models.}

\FullConference{%
  *** 27th European Cosmic Ray Symposium - ECRS ***\\
  *** 25-29 July 2022 ***\\
  *** Nijmegen, the Netherlands ***
}

\begin{document}
\maketitle

\section{Introduction}

High-energy cosmic rays are observed indirectly through the extensive air showers (EAS) they initiate in the Earth's atmosphere. The muon content of an EAS, together with a measure of the electromagnetic shower component, can be used to estimate the mass and energy of the primary cosmic ray. The interpretation of such measurements in terms of properties of the primary nucleus relies on detailed simulations of the EAS development. 
However, systematic differences between muon measurements and predictions from simulations have been reported by a number of air-shower experiments~\cite{EAS-MSU:2019kmv}, preventing an unambiguous determination of the mass composition. These differences are expected to arise as a result of shortcomings in the description of energetic hadronic interactions, for which effective models tuned to accelerator data are used. Experiments can test and constrain hadronic interaction models by measuring EAS under different conditions.

With its combination of a surface air-shower array, IceTop, and a large-volume submerged in-ice detector, IceCube, the IceCube Neutrino Observatory can probe the muon component of EAS in two different energy regimes. The density of muons at the surface, dominated by GeV muons, has been measured at large lateral distance from the shower axis with IceTop for primary energies between \SI{2.5}{\peta\eV} and \SI{120}{\peta\eV}~\cite{IceCube:2022yap}. In this article, the average multiplicity of high-energy muons in the shower accessible with the IceCube detector, referred to as TeV muons, is determined. The measurement of muons in different energy regimes can provide unique constraints on muon production models~\cite{Riehn:2019jet}. A first step in this direction, indicating inconsistencies in the models \sibyllpre{}~\cite{Ahn:2009wx}, \qgsjet{}~\cite{Ostapchenko:2010vb}, and \epos{}~\cite{Pierog:2013ria}, was presented in \refref{IceCube:2021ixw}.

\section{Cosmic rays with IceTop and IceCube}

The IceCube Neutrino Observatory is a multi-purpose particle detector located at the geographical South Pole, consisting out of the IceTop~\cite{IceCube:2012nn} and IceCube~\cite{IceCube:2016zyt} detectors.

IceTop is a square-kilometer air-shower array deployed at the surface, which has an altitude of \SI{2835}{\m} corresponding to an atmospheric depth of about \SI{690}{\g\per\cm\squared}. It consists of 81 stations on a triangular grid with a horizontal spacing of about \SI{125}{\m}. Every station comprises two ice-Cherenkov tanks, containing two Digital Optical Modules (DOMs) each, which detect the light produced by shower particles penetrating the tanks. IceTop performs optimally for EAS from cosmic rays in the primary energy range of \SI{1}{\peta\eV} to \SI{1}{\exa\eV}. Due to its high elevation, the array sits close to the depth of shower maximum, and the IceTop signals are dominated by the electromagnetic shower component. The signal contribution from low-energy ($\sim$\si{\GeV}) muons becomes visible over the electromagnetic contribution at several hundred meters from the shower axis.

The in-ice detector is located at depths between \SI{1.5}{\km} and \SI{2.5}{\km} below the surface. Over 5000 DOMs are deployed on vertical strings following approximately the same pattern as the IceTop array, with a vertical spacing of \SI{17}{\m}. While IceCube is primarily designed to detect the charged particles originating from neutrino interactions in the ice, its trigger rate is dominated by atmospheric muons with energies of several hundred GeV penetrating the ice. When the geometry allows it, an EAS triggering IceTop can be accompanied by a bundle of such high-energy muons leaving behind a signal in IceCube. 

\section{TeV muon multiplicity analysis}

The idea for the TeV muon multiplicity analysis is to examine vertical EAS which trigger IceTop and have a coincident muon bundle in IceCube. The IceTop signals are used to reconstruct the direction of the shower and to obtain a proxy for the primary cosmic-ray energy. The in-ice signals are used to reconstruct the energy loss profile of the muon bundle throughout the detector. The resulting information is combined in a neural network which predicts the primary cosmic ray energy \Enul{}, as well as the multiplicity of muons with an energy above \SI{500}{\giga\eV} in the shower \Nmu{}. The energy threshold of \SI{500}{\giga\eV} is chosen because most vertical muons with this energy are expected to reach IceCube. \Nmu{} is defined as the number of such muons present in the shower at the surface. Using the event-by-event neural network predictions, the average multiplicity $\langle N_\mu \rangle$ is obtained in bins of \Enul{}. Corrections derived from simulation are applied to deal with biases resulting from imperfections in the neural-network reconstruction.

\subsection{Air-shower and energy-loss reconstructions}\label{sec:recos}

An air-shower reconstruction is applied to events triggering IceTop. The tank signals are calibrated in units of vertical equivalent muons (VEM) and several algorithms selecting hits that are likely related to a single shower are applied~\cite{IceCube:2012nn}. The resulting signals are used in a maximum-likelihood reconstruction in which the core position and direction of the shower axis are reconstructed by means of fitting the signal times and charges as a function of radial distance to the shower axis, as described in \refref{IceCube:2013ftu}. 
The expected signal strength as a function of radial distance is defined by two free parameters, one of which is the expected signal strength at a reference distance of \SI{125}{\m}, referred to as \Sone{}.
Snow accumulation on top of the tanks as a result of the environmental conditions at the South Pole causes a fraction of the shower particles to be absorbed; this is taken into account in an approximate way during reconstruction.

When signals that could be causally related to the same air shower are present in IceCube, further processing is applied as described in \refref{IceCube:2019hmk}. The shower axis resulting from the IceTop reconstruction is used to select signals likely caused by the high-energy muon bundle. The shower axis is also used as a seed track for a reconstruction of the energy loss of the muon bundle throughout the detector. Based on the charge and time information of the signals, the deposited energy is reconstructed in segments of length \SI{20}{\m} with the algorithm described in \refref{IceCube:2013dkx}.

\subsection{Event selection}\label{sec:event}

The event selection for this analysis aims to obtain well-reconstructed air-shower events that are coincident between IceTop and IceCube. To this end, the quality cuts from \refref{IceCube:2019hmk} are applied. A number of IceTop cuts aims to select air showers that have their core contained within the boundary of IceTop. Combined with cuts requiring a minimal number of stations left after cleaning and a minimal charge of \SI{6}{VEM} present in the event after the snow correction, a subset of events remains where the air-shower reconstruction has an excellent performance. The IceCube selection requires a minimum number of hits to be present after cleaning and applies various cuts related to the quality of the energy loss reconstruction.

The combination of the  IceTop and IceCube selection limits the zenith angle range of showers to $\cos\theta \gtrsim 0.85$ ($\theta \lesssim 32^\circ$). A further cut is applied to the reconstructed zenith angle to select near-vertical EAS with $\cos\theta > 0.95$ ($\theta \lesssim 18^\circ$). This avoids possible complications from zenith-dependent effects expected in both the muon production in the atmosphere and the propagation toward the detector.

The selection reaches full efficiency for primary nuclei of all types around \SI{2.5}{\peta\eV}. Above this threshold energy, the resolution of the core position is better than \SI{12}{\m} and the angular resolution is below $1^\circ$~\cite{IceCube:2019hmk}. Simulations show that the shower size \Sone{} is strongly correlated with the primary energy with a small dependence on the primary mass. Similarly, the average reconstructed energy loss in IceCube is found to be related to the multiplicity of TeV muons in the shower.

\subsection{Neural network model}

\begin{figure}
    \centering
    \includegraphics[trim=0 2em 0 1.5em, clip, width=0.45\textwidth]{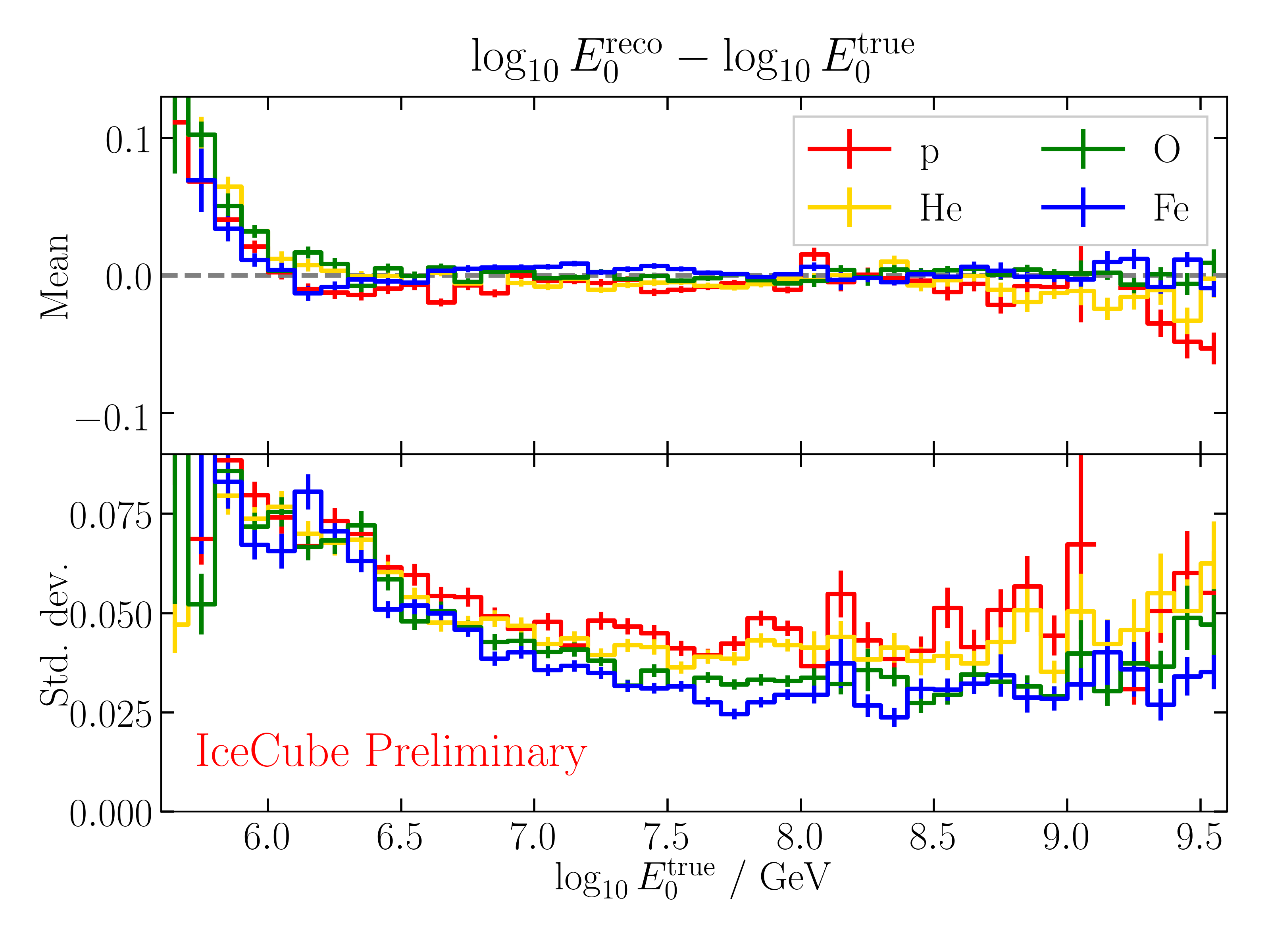}\includegraphics[trim=0 2em 0 1.5em, clip, width=0.45\textwidth]{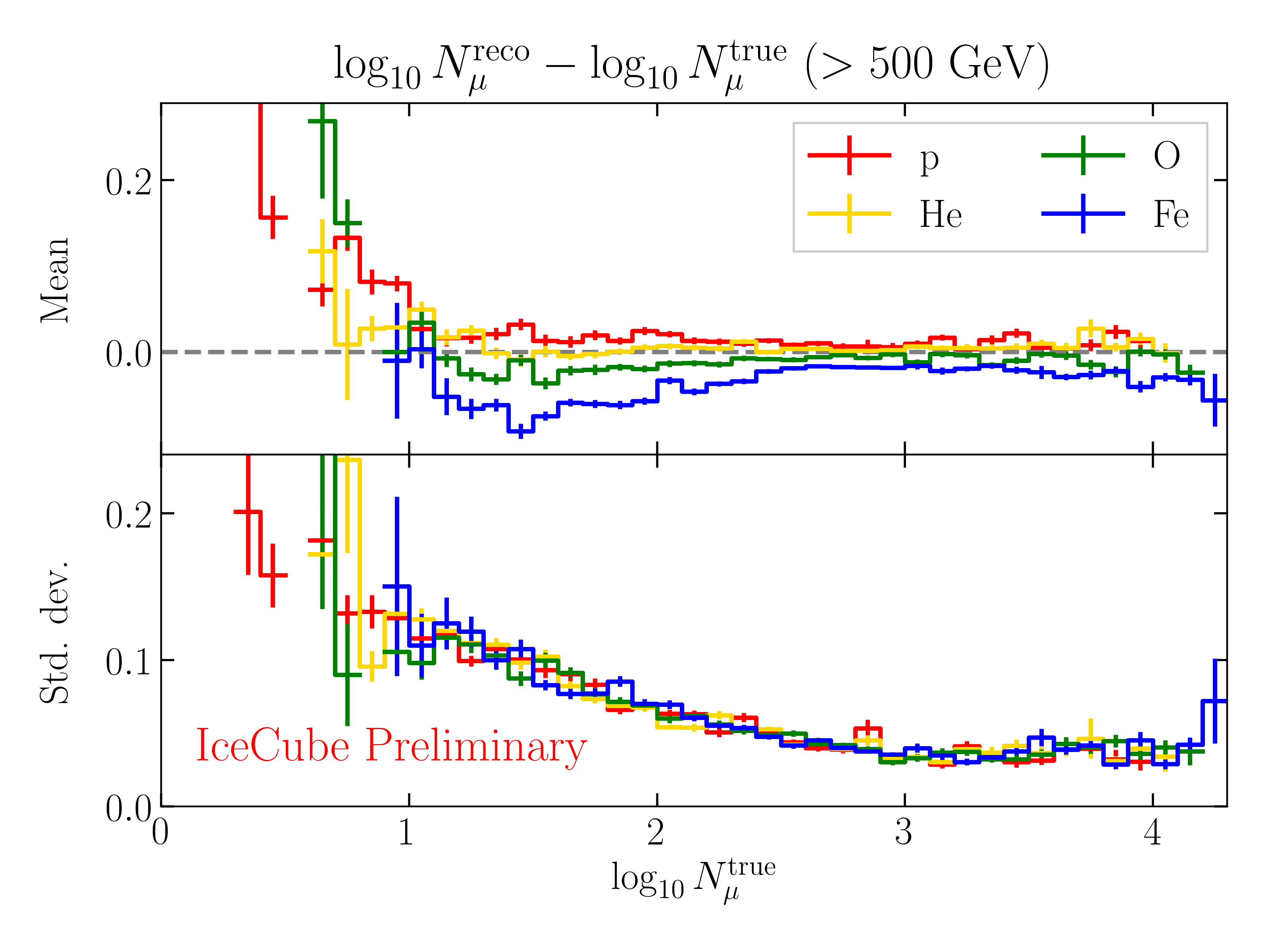}
    \caption{Performance of the NN primary energy (left) and muon multiplicity (right) reconstruction. Shown are the bias and resolution, defined as the mean and standard deviation of the difference between the true and reconstructed quantities in logarithmic scale. The x\nobreakdash-axis depicts the true quantity.}
    \label{fig:performance}
\end{figure}

A neural network is trained to predict the primary energy \Enul{} and multiplicity of muons above \SI{500}{\GeV} \Nmu{} in the shower at surface level, based on the reconstructions of \refsec{sec:recos}: shower size \Sone{}, reconstructed zenith angle \tta{}, and the reconstructed energy loss profile. The latter will be the most important input toward \Nmu{}, while \Sone{} will be most important for \Enul{}. The addition of \tta{} may help to cover small zenith-dependent effects in either of the previous two relations.

The reconstructed energy loss profile is transformed to a vector where each entry corresponds to a reconstructed energy loss value in a \SI{20}{\m} segment. Based on the event geometry, the vector is padded with zeros to obtain a vector of fixed length, in such a way that each entry corresponds to a particular slanth depth of the muon bundle in the ice. Events with less than three non-zero entries are discarded.

The energy loss vector forms the input for a recurrent neural network (RNN) layer, ideal for sequential data. The output of the RNN layer is concatenated with \Sone{} and \tta{}. Through a series of dense layers, the neural network finally returns a value for \Enul{} and \Nmu{}.

The neural network is trained on Monte Carlo (MC) simulation performed with CORSIKA v73700~\cite{Heck:1998vt} using an April South Pole atmosphere 
and \sibyllpre{} as the high-energy hadronic interaction model. It has undergone a full detector simulation, after which the event selection of \refsec{sec:event} has been applied. The dataset includes four primary types (p, He, O, Fe). The \Enul{}-range covered by the dataset after event selection runs approximately from \SI{300}{\tera\eV} to \SI{4}{\exa\eV}, corresponding to an \Nmu{}-range of approximately 1 to \num{20000}. The loss function is a simple combination of a mean-squared error loss for both targets (in $\log_{10}$).

The performance of the neural network, derived from a separate dataset not included in the training, is shown in \reffig{fig:performance}. In general, good performance is found above the IceTop threshold energy of $\log_{10} E_0 / \si{\GeV} = 6.4$, corresponding to a $\log_{10} N_\mu$ of about 1.3 or 1.7 for proton and iron, respectively. The \Enul{}-reconstruction has an excellent resolution, performing somewhat better for heavier primaries, and shows little bias in the energy range of interest. The resolution of the \Nmu{}-reconstruction improves with the multiplicity up to around 1000 muons in a nearly mass-independent way. This reconstruction shows, however, clear mass-dependent biases.

\subsection{MC Correction}\label{sec:correction}

\begin{figure}
    \centering
    \includegraphics[trim=0 2em 0 1.5em, clip, width=0.427\textwidth]{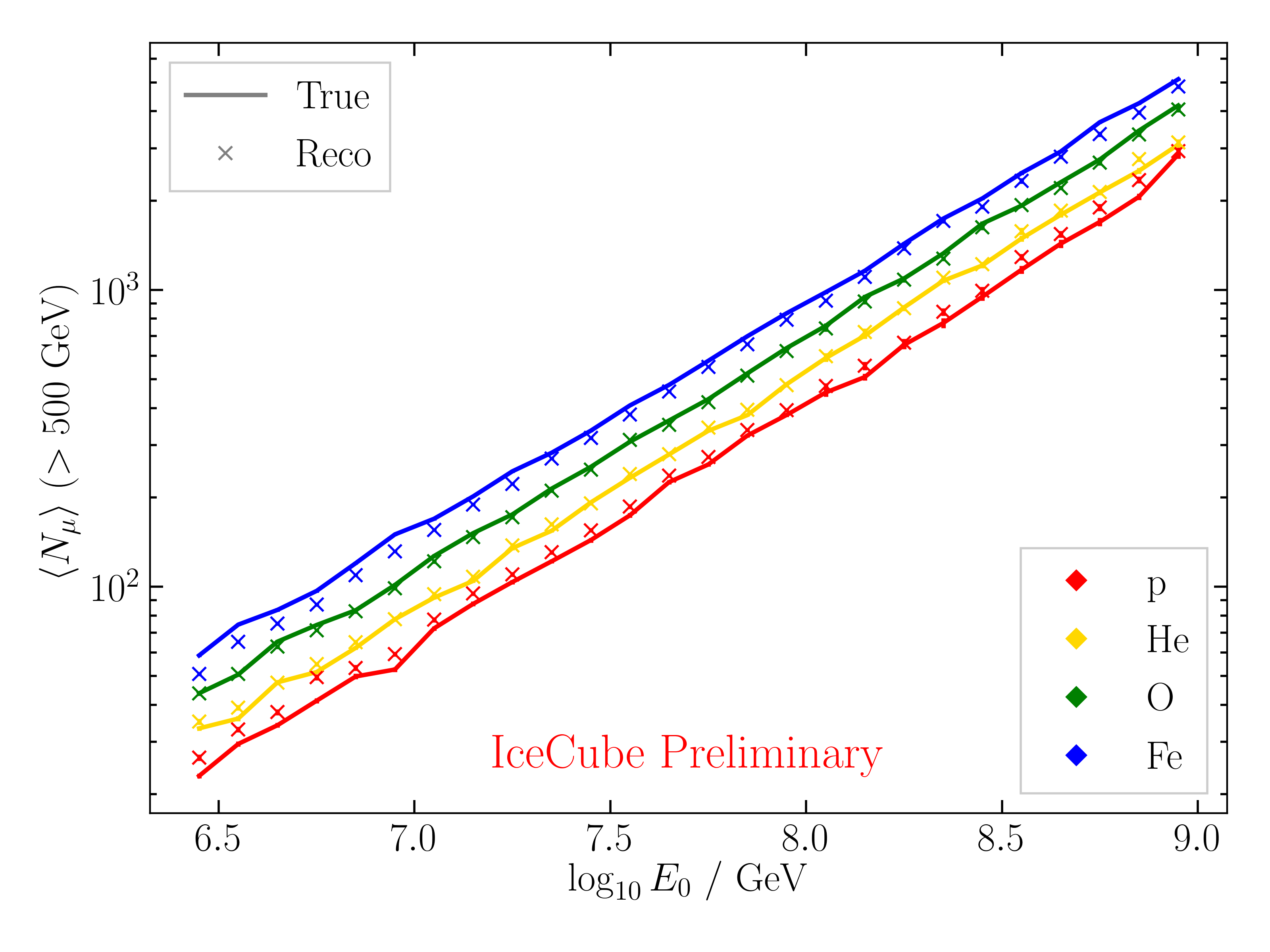}\includegraphics[trim=0 2em 0 1.5em, clip, width=0.45\textwidth]{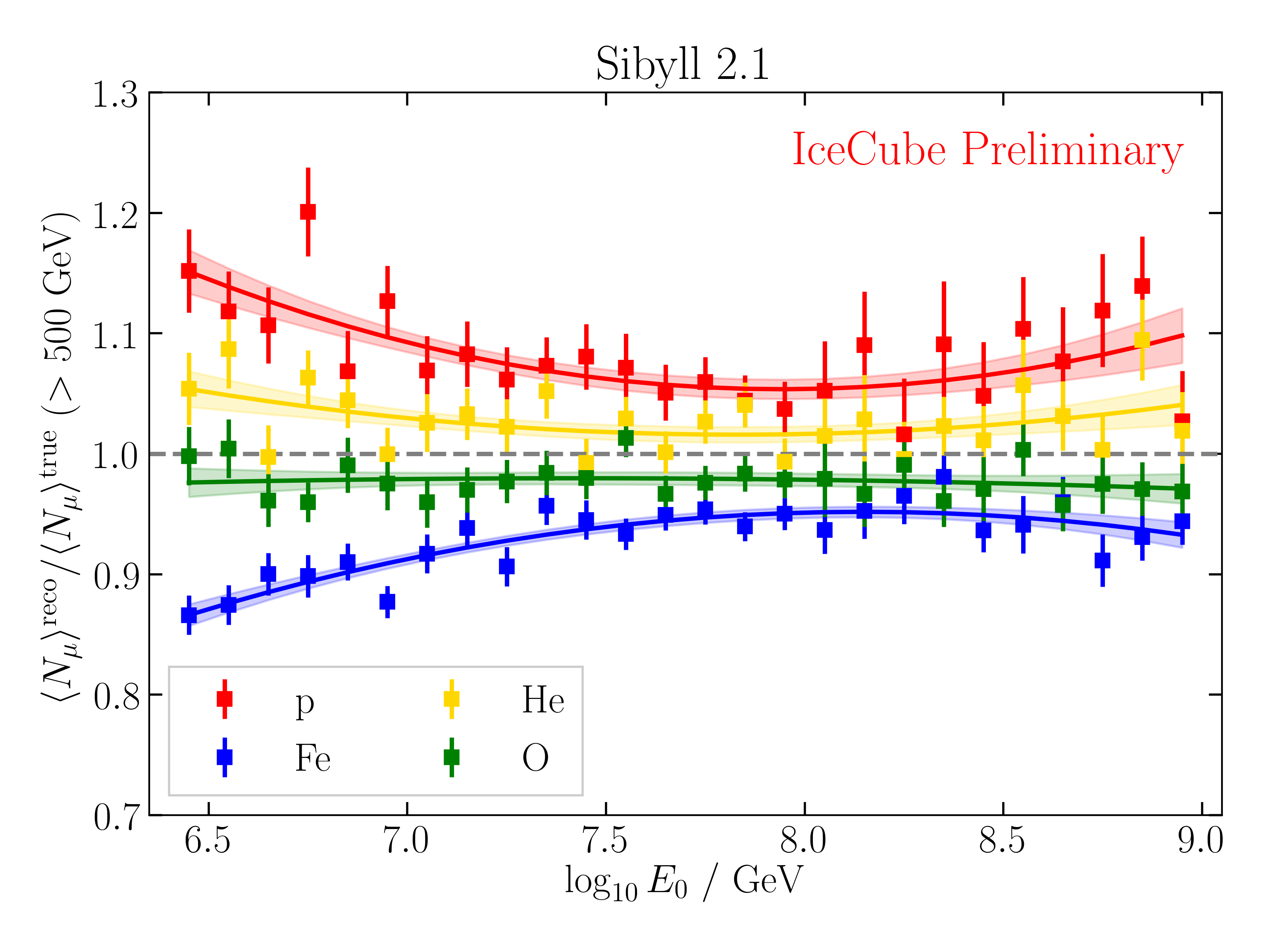}
    \caption{Average muon multiplicity ($> \SI{500}{\GeV}$) as a function of primary energy in simulation obtained using the neural network predictions of \Nmu{} and \Enul{} compared to the true values. The right plot shows the ratio of the reconstructed and true values, together with quadratic fits to be used as correction factors.}
    \label{fig:bias}
\end{figure}

The goal of the neural network reconstruction is to find the average muon multiplicity as a function of primary energy. \reffig{fig:bias} (left) shows the average reconstructed \Nmu{} in bins of reconstructed \Enul{} derived from MC, compared to the true \Nmu{} in bins of true \Enul{}. The results derived from the neural network approximate the general relation between \Nmu{} and \Enul{} and the differences between primaries, but some systematic deviations from the true values can be observed. This is demonstrated in detail in the ratio plot (right), showing clear mass-dependent biases up to $\sim 15\%$. These deviations result from imperfections in the \Nmu{}-reconstruction as well as the \Enul{}-reconstruction due to bin migration. This effect should be taken into account when applying the reconstruction to experimental data. To this end, the ratios between the true and reconstructed \Nav{} are fitted with quadratic functions, as shown in the right panel of \reffig{fig:bias}. These fits can be used as multiplicative correction factors to correct for the reconstruction bias.

Given that the mass composition is not known, it is not clear which (combination of) correction factor(s) should be applied to experimental data. However, the fact that a measurement of \Nmu{} itself has composition information, and that the correction factors vary smoothly with the logarithmic mass $\ln A$ (p, He, O, Fe are approximately equidistant in $\ln A$, as are the correction factors in \reffig{fig:bias}), suggests an iterative procedure to deal with this issue. A common way of expressing muon measurements is by comparing them to the predictions of pure proton and iron MC, $\langle N_\mu \rangle_\mathrm{p}$ and $\langle N_\mu \rangle_\mathrm{Fe}$, as
\begin{equation}\label{eq:z}
    z = \frac{\ln \langle N_\mu \rangle - \ln \langle N_\mu\rangle_\mathrm{p}}{\ln \langle N_\mu\rangle_\mathrm{Fe} - \ln \langle N_\mu\rangle_\mathrm{p}},
\end{equation}
the so-called ``z-scale''~\cite{EAS-MSU:2019kmv}. The Heitler-Matthews model and the superposition approximation imply that $z \approx \ln A / \ln 56$~\cite{Matthews:2005sd}, so that a measurement of \Nav{} implies a certain average composition. The proton and iron correction factors can then be combined to an effective correction factor $\mathcal{C}_\mathrm{eff}$ corresponding to this composition as
\begin{equation}\label{eq:correction}
    \mathcal{C}_\mathrm{eff}(\ln A) = \mathcal{C}_\mathrm{p} + \frac{\mathcal{C}_\mathrm{Fe} - \mathcal{C}_\mathrm{p}}{\ln 56} \ln A.
\end{equation}
New estimates for \Nav{} and $\mathcal{C}_\mathrm{eff}$ can then be obtained in an iterative way until the difference in \Nav{} between two subsequent steps becomes negligiblly small. In this way, results are obtained without any prior composition assumption. The statistical uncertainty on the final correction factor, i.e. the uncertainty bands on $\mathcal{C}_\mathrm{p}$ and $\mathcal{C}_\mathrm{Fe}$ from \reffig{fig:bias} propagated through \refeq{eq:correction}, will be taken into account as a systematic uncertainty on the final result.

\begin{wrapfigure}{}{0.45\textwidth}
    \centering
    \includegraphics[trim=0 2em 0 1.5em, clip, width=0.432\textwidth]{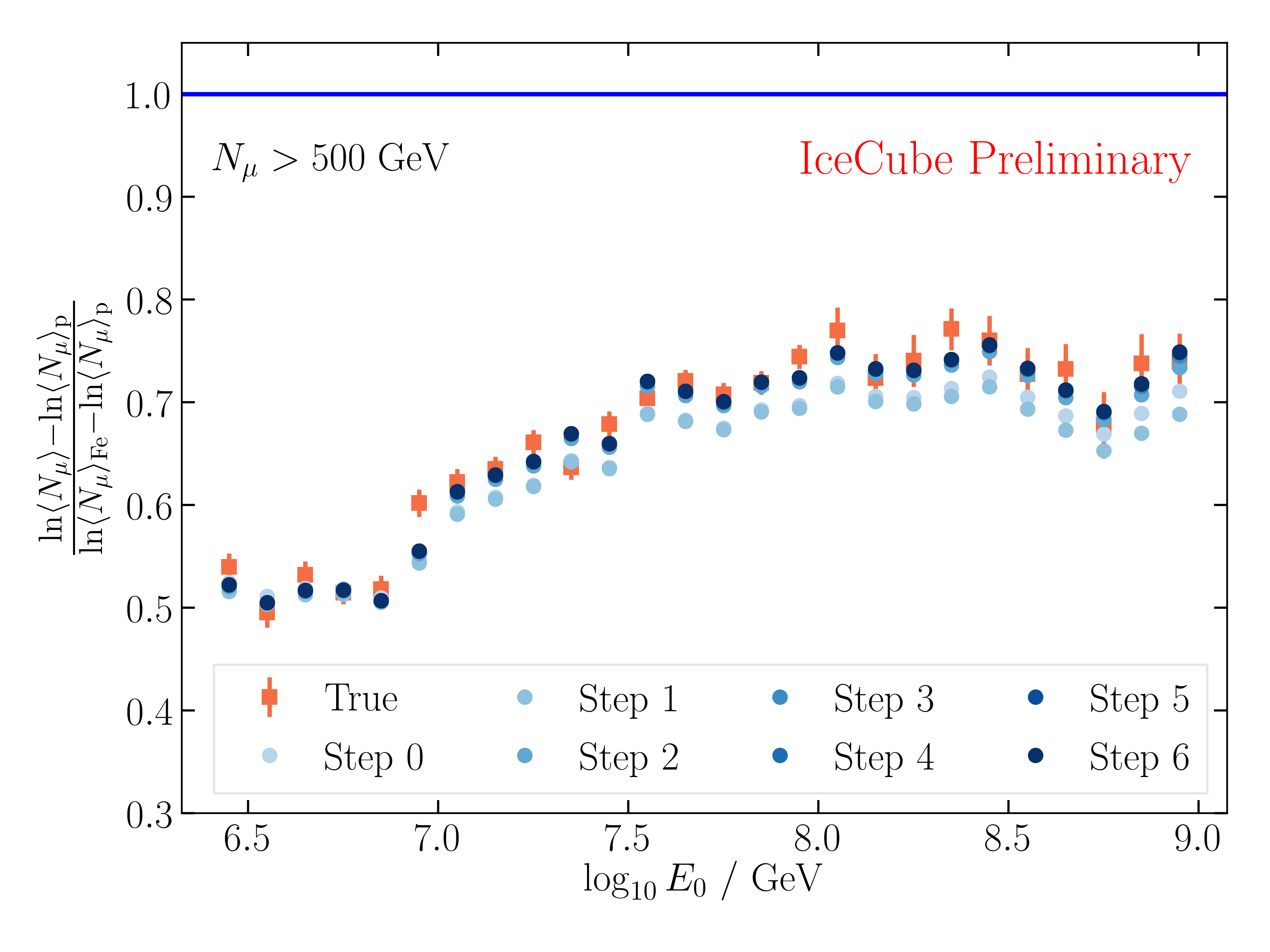}
    \caption{Example of the iterative correction procedure applied to MC weighted
according to H4a. After several iteration steps, the reconstructed \Nav{} values (blue) converge to values consistent with the MC truth (orange).}
    \label{fig:H4a}
\end{wrapfigure}

An example of this iterative procedure for simulation weighted according to the H4a composition model is shown in \reffig{fig:H4a}. After several steps, the reconstructed values are in agreement with the true values. Tests on a large variety of composition assumptions have shown that this method works generally.

The correction factors shown in this section were derived from \sibyllpre{} MC. By applying the neural network to MC based on different hadronic interaction models, other correction factors can be derived, and results can be obtained from experimental data under the assumption that the different models give a correct description of reality.

\section{Results}

The method discussed in the previous section is applied to 10\% of data obtained between May 2012 and May 2013. The average atmosphere and snow coverage of this period agree well with the simulated April atmosphere and the simulated snow coverage corresponding to in-situ measurements from October 2012~\cite{DeRidder:2019}.

\begin{figure}
    \centering
    \includegraphics[trim=0 1.7em 0 1.4em, clip, width=0.45\textwidth]{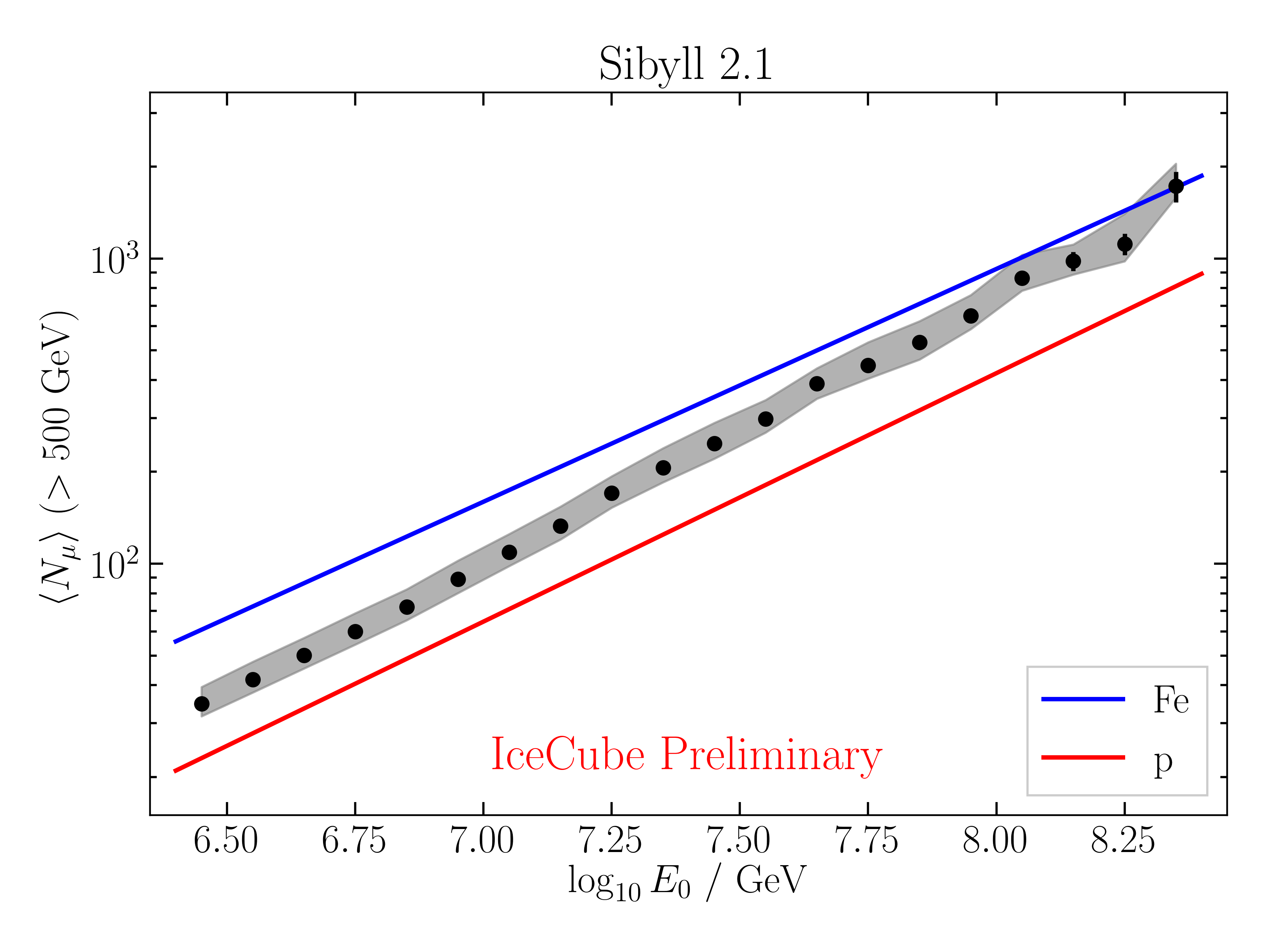}\includegraphics[trim=0 1.7em 0 1.4em, clip, width=0.45\textwidth]{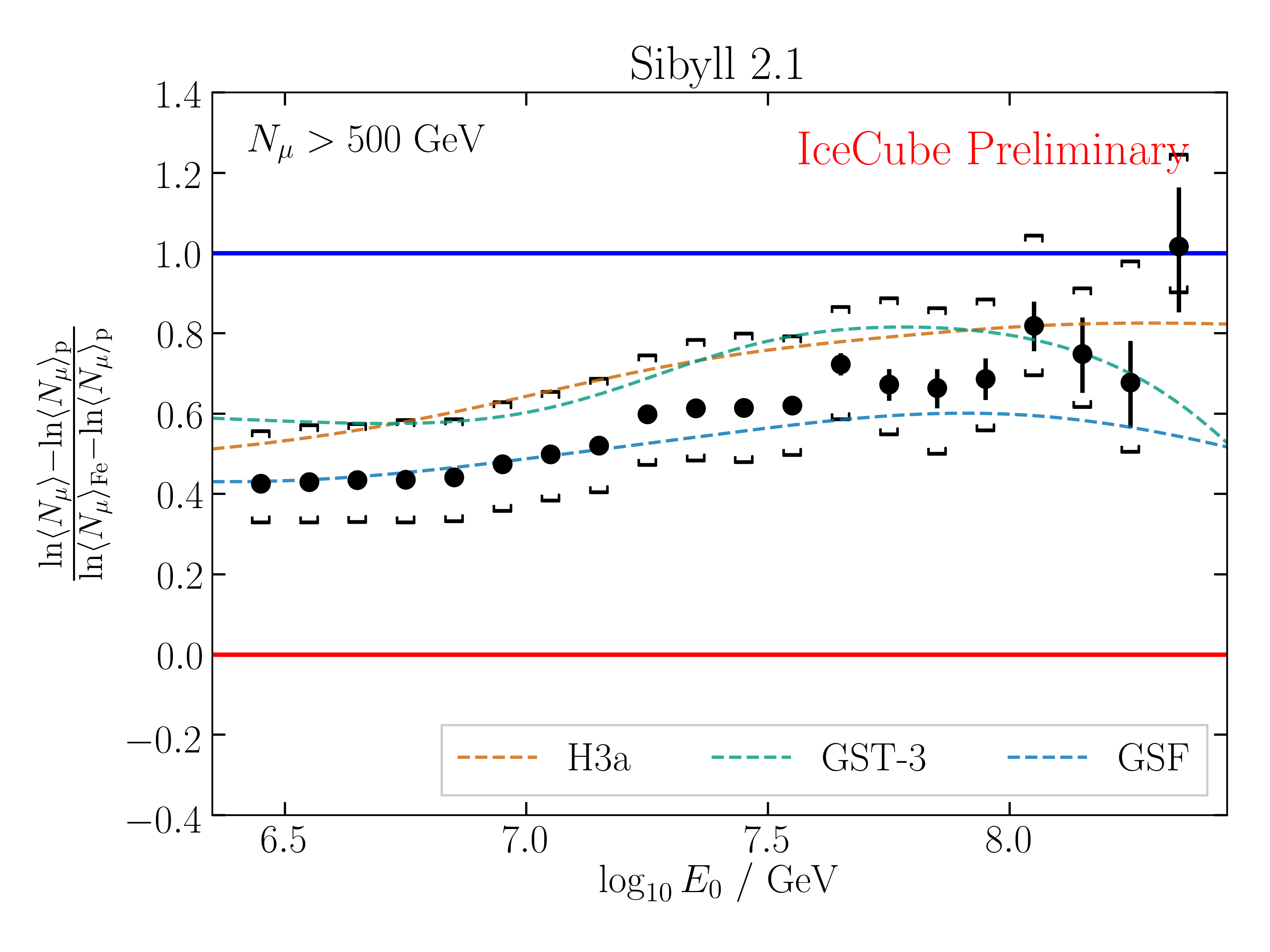}

    \includegraphics[trim=0 1.7em 0 1.4em, clip, width=0.45\textwidth]{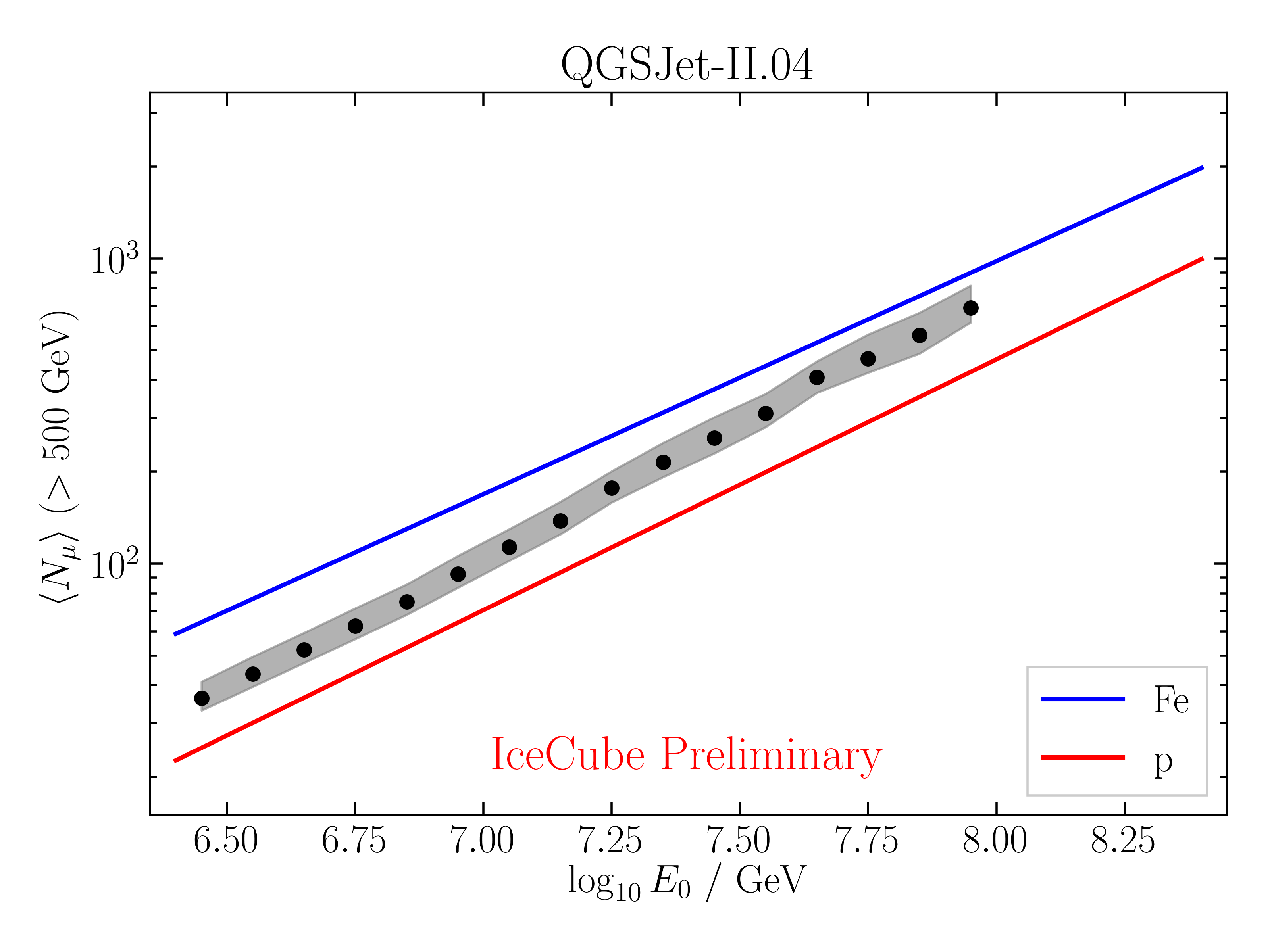}\includegraphics[trim=0 1.7em 0 1.4em, clip, width=0.45\textwidth]{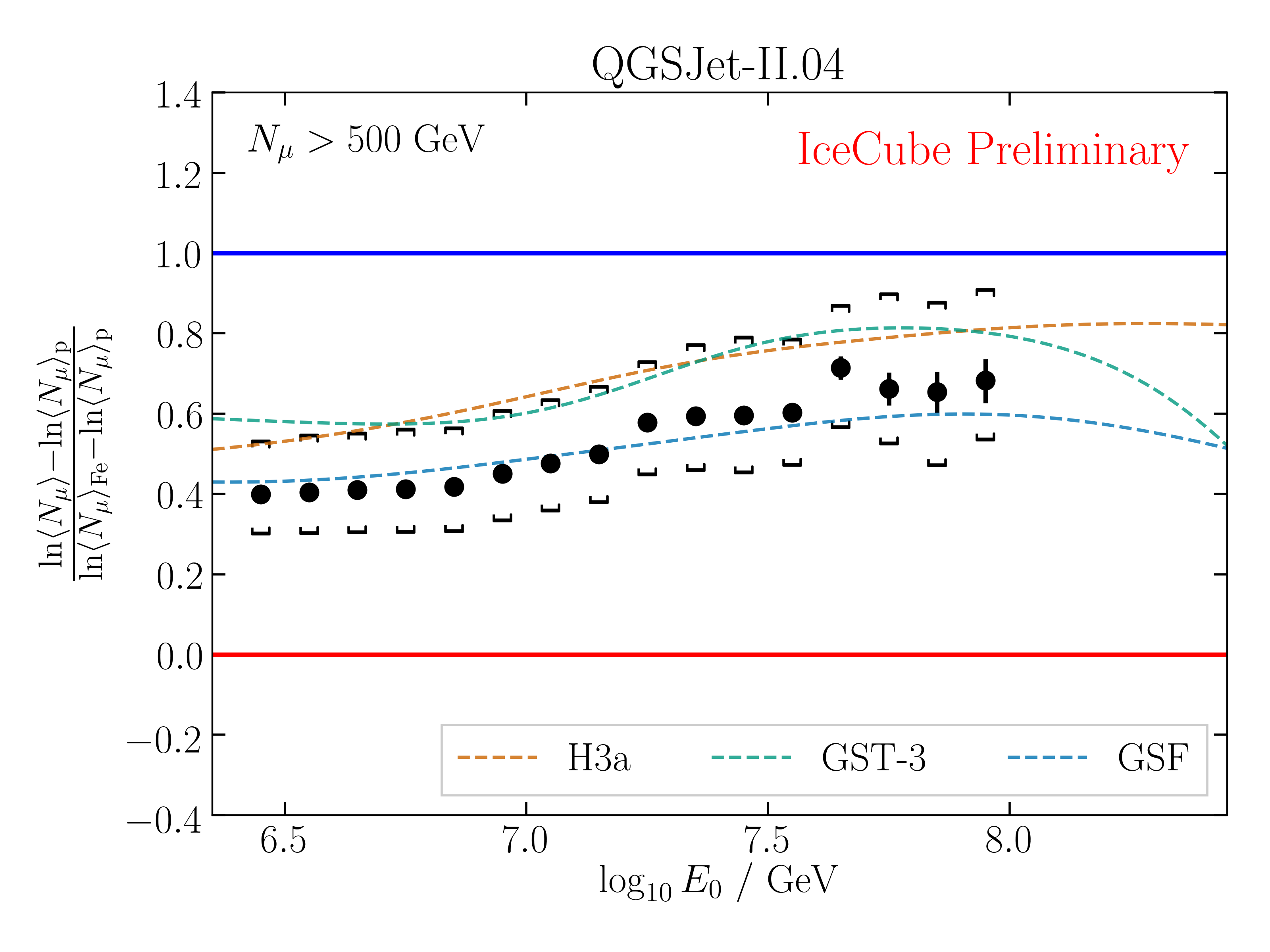}

    \includegraphics[trim=0 1.7em 0 1.4em, clip, width=0.45\textwidth]{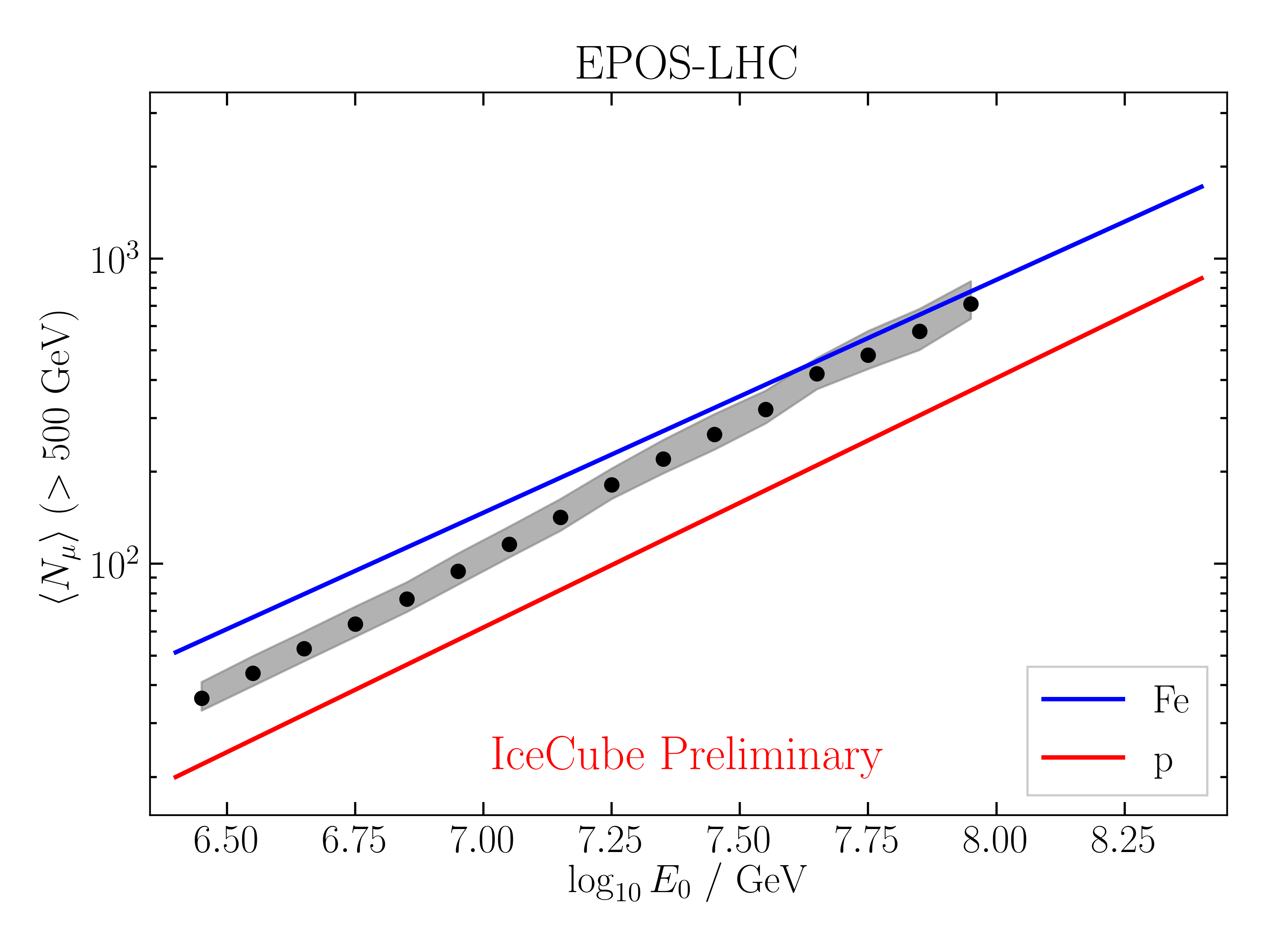}\includegraphics[trim=0 1.7em 0 1.4em, clip, width=0.45\textwidth]{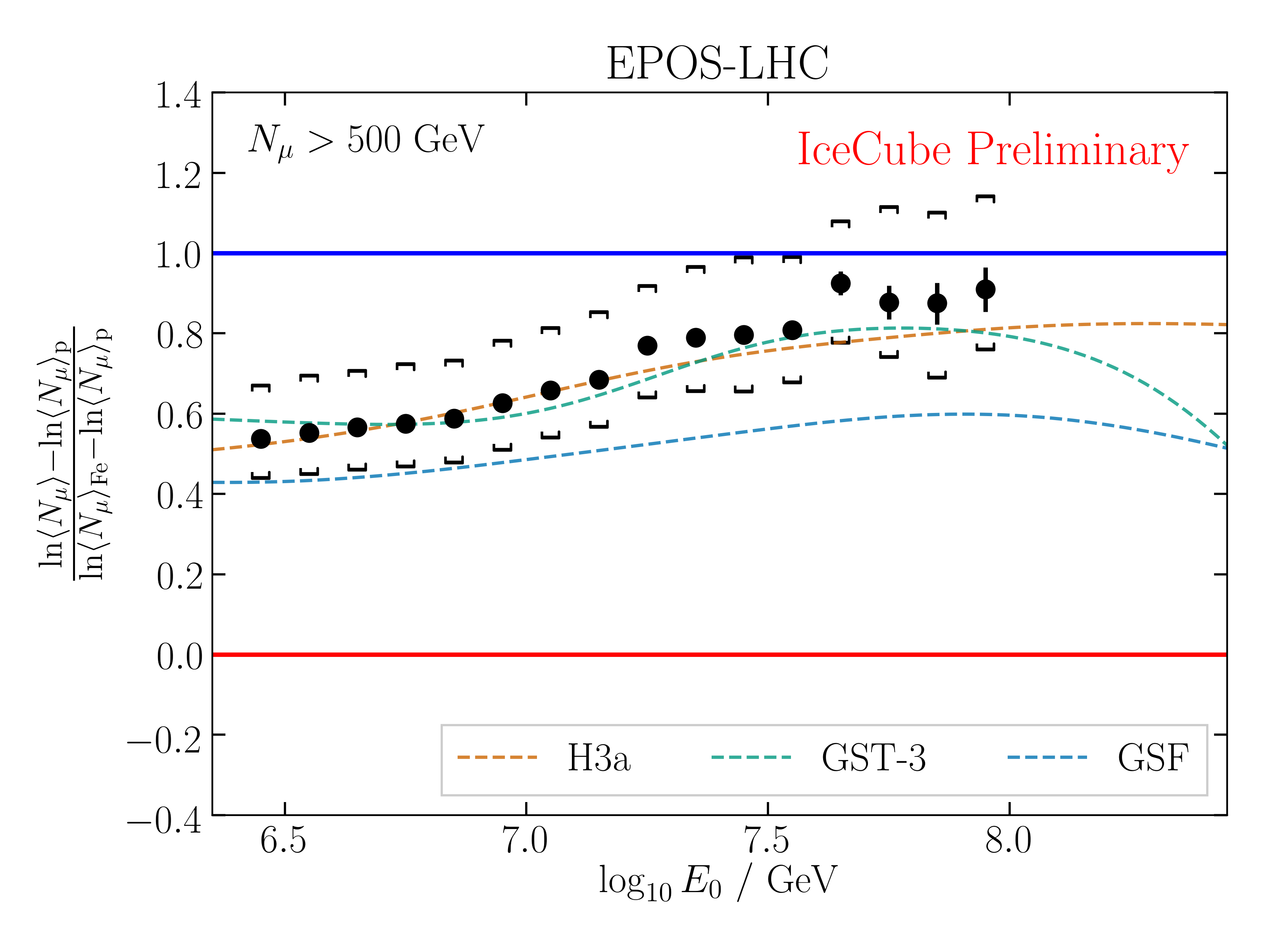}

    \caption{Average multiplicity of muons with energy larger than 500 GeV as a function of primary energy for near-vertical EAS, obtained from experimental data under the assumption of different hadronic interaction models. The figures on the right show the corresponding $z$-values (\refeq{eq:z}), with expectations from composition models for comparison. Error bars indicate the statistical uncertainty, bands/brackets indicate the total systematic uncertainty.}
    \label{fig:results}
\end{figure}

Results are obtained based on the correction factors derived from the \sibyllpre{} dataset, as well as from two more datasets based on \qgsjet{} and \epos{}, which are limited to \SI{100}{\peta\eV} in energy. They are shown in \reffig{fig:results}. The variation in \Nav{} obtained assuming different models is $\lesssim 8\%$.

Several systematic uncertainties can influence the analysis. The detector systematics included here are the same as in \refref{IceCube:2019hmk}: the snow correction for IceTop, the calibration of the VEM unit, and the in-ice light-yield, related to uncertainties in the ice model and DOM efficiency. All of these uncertainties are added in quadrature, together with the uncertainty on the correction factors (\refsec{sec:correction}).

The \Nav{} results in \reffig{fig:results} are compared to expectations for proton and iron, obtained from dedicated simulations using the corresponding models. The results are bracketed by proton and iron for all models. In the $z$-representation (\refeq{eq:z}), expectations from different composition models are also shown: H3a~\cite{Gaisser:2011klf}, GST~\cite{Gaisser:2013bla}, and GSF~\cite{Dembinski:2017zsh}. The results for \sibyllpre{} and \qgsjet{} are very similar in this representation and agree well with the expectations, especially from GSF. The result for \epos{} indicates a somewhat heavier composition, mainly because \epos{} predicts less TeV muons, but is still in agreement with expectations.

\section{Conclusion and Outlook}

We have presented an analysis of the multiplicity of muons with energy above \SI{500}{\GeV} in near-vertical EAS observed in IceTop and IceCube. Results were obtained assuming different hadronic interaction models, covering a primary energy range from \SI{2.5}{\peta\eV} to \SI{250}{\peta\eV} for \sibyllpre{} and up to \SI{100}{\peta\eV} for \qgsjet{} and \epos{}. In all cases, results are found to be in agreement with expectations from recent composition models.

It is interesting to compare the composition interpretation of the TeV muon measurement obtained here to the measurement of the GeV muon density in IceTop presented in \refref{IceCube:2022yap}. While consistent results are found for the GeV and TeV muons when assuming \sibyllpre{}, a tension is observed for \qgsjet{} and \epos{}, where the GeV muons indicate a lighter composition.

The analysis can be improved in several ways in the future, e.g. through a larger experimental dataset, an increased zenith range, and reduced systematic uncertainties. The phase space and accuracy of muon measurements are furthermore expected to increase with planned detector upgrades, such as the enhancement of surface instrumentation~\cite{Haungs:2019ylq} and the larger size of IceCube-Gen2~\cite{IceCube-Gen2:2020qha}.

\bibliographystyle{ICRC}
\bibliography{references}

\providecommand{\href}[2]{#2}\begingroup\raggedright\setstretch{0.01}\begin{thebibliography}{10}

\bibitem{EAS-MSU:2019kmv}
H.~P. Dembinski {\em et~al.}
  \href{http://dx.doi.org/10.1051/epjconf/201921002004}{{\em EPJ Web Conf.}
  {\bfseries 210} (2019) 02004}.

\bibitem{IceCube:2022yap}
{\bfseries IceCube} Collaboration, R.~Abbasi {\em et~al.}
  \href{http://dx.doi.org/10.1103/PhysRevD.106.032010}{{\em Phys. Rev. D}
  {\bfseries 106} no.~3, (2022) 032010}.

\bibitem{Riehn:2019jet}
F.~Riehn, R.~Engel, A.~Fedynitch, T.~K. Gaisser, and T.~Stanev
  \href{http://dx.doi.org/10.1103/PhysRevD.102.063002}{{\em Phys. Rev. D}
  {\bfseries 102} no.~6, (2020) 063002}.

\bibitem{Ahn:2009wx}
E.-J. Ahn, R.~Engel, T.~K. Gaisser, P.~Lipari, and T.~Stanev
  \href{http://dx.doi.org/10.1103/PhysRevD.80.094003}{{\em Phys. Rev. D}
  {\bfseries 80} (2009) 094003}.

\bibitem{Ostapchenko:2010vb}
S.~Ostapchenko \href{http://dx.doi.org/10.1103/PhysRevD.83.014018}{{\em Phys.
  Rev. D} {\bfseries 83} (2011) 014018}.

\bibitem{Pierog:2013ria}
T.~Pierog, I.~Karpenko, J.~M. Katzy, E.~Yatsenko, and K.~Werner
  \href{http://dx.doi.org/10.1103/PhysRevC.92.034906}{{\em Phys. Rev. C}
  {\bfseries 92} no.~3, (2015) 034906}.

\bibitem{IceCube:2021ixw}
{\bfseries IceCube} Collaboration, S.~Verpoest {\em et~al.}
  \href{http://dx.doi.org/10.22323/1.395.0357}{{\em PoS} {\bfseries ICRC2021}
  (2021) 357}.

\bibitem{IceCube:2012nn}
{\bfseries IceCube} Collaboration, R.~Abbasi {\em et~al.}
  \href{http://dx.doi.org/10.1016/j.nima.2012.10.067}{{\em Nucl. Instrum. Meth.
  A} {\bfseries 700} (2013) 188--220}.

\bibitem{IceCube:2016zyt}
{\bfseries IceCube} Collaboration, M.~G. Aartsen {\em et~al.}
  \href{http://dx.doi.org/10.1088/1748-0221/12/03/P03012}{{\em JINST}
  {\bfseries 12} no.~03, (2017) P03012}.

\bibitem{IceCube:2013ftu}
{\bfseries IceCube} Collaboration, M.~G. Aartsen {\em et~al.}
  \href{http://dx.doi.org/10.1103/PhysRevD.88.042004}{{\em Phys. Rev. D}
  {\bfseries 88} no.~4, (2013) 042004}.

\bibitem{IceCube:2019hmk}
{\bfseries IceCube} Collaboration, M.~G. Aartsen {\em et~al.}
  \href{http://dx.doi.org/10.1103/PhysRevD.100.082002}{{\em Phys. Rev. D}
  {\bfseries 100} no.~8, (2019) 082002}.

\bibitem{IceCube:2013dkx}
{\bfseries IceCube} Collaboration, M.~G. Aartsen {\em et~al.}
  \href{http://dx.doi.org/10.1088/1748-0221/9/03/P03009}{{\em JINST} {\bfseries
  9} (2014) P03009}.

\bibitem{Heck:1998vt}
D.~Heck {\em et~al.} {\em Wissenschaftliche Berichte, Forschungszentrum
  Karlsruhe} (1998) .

\bibitem{Matthews:2005sd}
J.~Matthews \href{http://dx.doi.org/10.1016/j.astropartphys.2004.09.003}{{\em
  Astropart. Phys.} {\bfseries 22} (2005) 387--397}.

\bibitem{DeRidder:2019}
S.~De~Ridder, {\em {Sensitivity of IceCube cosmic ray measurements to the
  hadronic interaction models}}.
\newblock PhD thesis, {Ghent University, Belgium}, {2019}.

\bibitem{Gaisser:2011klf}
T.~K. Gaisser \href{http://dx.doi.org/10.1016/j.astropartphys.2012.02.010}{{\em
  Astropart. Phys.} {\bfseries 35} (2012) 801--806}.

\bibitem{Gaisser:2013bla}
T.~K. Gaisser, T.~Stanev, and S.~Tilav
  \href{http://dx.doi.org/10.1007/s11467-013-0319-7}{{\em Front. Phys.
  (Beijing)} {\bfseries 8} (2013) 748--758}.

\bibitem{Dembinski:2017zsh}
H.~P. Dembinski {\em et~al.} \href{http://dx.doi.org/10.22323/1.301.0533}{{\em
  PoS} {\bfseries ICRC2017} (2018) 533}.

\bibitem{Haungs:2019ylq}
{\bfseries IceCube} Collaboration, A.~Haungs
  \href{http://dx.doi.org/10.1051/epjconf/201921006009}{{\em EPJ Web Conf.}
  {\bfseries 210} (2019) 06009}.

\bibitem{IceCube-Gen2:2020qha}
{\bfseries IceCube-Gen2} Collaboration, M.~G. Aartsen {\em et~al.}
  \href{http://dx.doi.org/10.1088/1361-6471/abbd48}{{\em J. Phys. G} {\bfseries
  48} no.~6, (2021) 060501}.

\end{thebibliography}\endgroup

\end{document}